# Looking for nebular He II emission south of the multiple massive-star system, HD 5980.

A. Sixtos[1],⋆ A. Wofford[1], A.A.C. Sander[2] and A. Peimbert[3]

[1]*Instituto de Astronomía, Universidad Nacional Autónoma de México, Unidad Académica en Ensenada, Km 103 Carr. Tijuana−Ensenada, Ensenada, B.C., C.P. 22860, México*

[2]*Zentrum für Astronomie der Universität Heidelberg, Astronomisches Rechen-Institut, Mönchhofstr. 12-14, 69120 Heidelberg*

[3]*Instituto de Astronomía, Universidad Nacional Autónoma de México, Apdo. Postal 70-264 Ciudad Universitaria, México*

**ABSTRACT**

The nebular He II $\lambda$1640 emission line is observed in star-forming galaxies out to large distances and can be used to constrain the properties of sources of He$^+$-ionizing photons. For this purpose, it is crucial to understand which are the main stellar sources of these photons. In some nearby metal-poor starburst galaxies, nebular He II $\lambda$4686 (optical-equivalent) is accompanied by a broad underlying component, which is generally attributed to formation in the winds of classical (He-burning) Wolf-Rayet (WR) stars, primarily of the WN subtype. In such cases, the origin of the nebular component has been proposed to be the escape of He$^+$-ionizing photons from the winds of the WN stars, at least partially. We use archival long-slit observations obtained with FORS1 on the *VLT* to look for nebular He II $\lambda$4686 emission south of the WN6h + WN6-7 close binary in HD 5980. We only find broad He II $\lambda$4686 emission, as far as ∼ 7.6 pc from the binary. A comparison with observations obtained with STIS on the *HST*, at a similar orbital phase shows that the FORS1 broad He II emission is likely contamination from the multiple-star system HD 5980. We use models to show that no significant He$^+$-ionizing flux is expected from the WN stars in HD 5980 and that when similar stars are present in a coeval stellar population, the O stars can be far greater emitters of He$^+$-ionizing radiation.

**Key words:** binaries: eclipsing – stars: individual (HD 5980) – stars: Wolf–Rayet – HII regions – Magellanic Clouds

## 1 INTRODUCTION.

### 1.1 Nebular He II from star-forming objects.

The presence of narrow (FWHM < 1000 km s$^{-1}$) nebular He II emission lines at 4686 Å (optical) and/or 1640 Å (ultraviolet, UV) in the integrated spectrum of a galaxy indicates the existence of a hard ionizing source, as photons of ≥ 54.4 eV ($\lambda$ ≤ 228 Å) are required to ionize He$^+$. It is the recombination of free electrons with He$^{2+}$ that produces the He II emission lines. Active galactic nuclei (AGN) are sources of hard radiation and produce luminous He II emission in galaxies (e.g., Shirazi & Brinchmann 2012; Saxena et al. 2020). However, nebular He II emission is also observed in star-forming (SF) galaxies without AGN signatures, particularly at subsolar metallicities (e.g., Shirazi & Brinchmann 2012; Senchyna et al. 2017; James et al. 2022), and as far as redshifts of $z$ ∼ 2−3, e.g. in galaxies of the VANDELS survey, which used VIMOS on the *Very Large Telescope* (*VLT*, Saxena et al. 2020). At even higher distances, in the Epoch of Reionization (EoR), tentative detections of He II emission from SF objects include: i) A1703-zd6 at $z$ ∼ 7, observed with MOSFIRE on Keck (Stark et al. 2015); ii) CR7 OB3 at $z$ ∼ 6.6, observed with XShooter on the *VLT* (Sobral et al. 2019); and iii) a galaxy at $z$ = 6.112, observed with NIRCam on the *James Webb Space Telescope (JWST)*, as part of the commissioning of the wide-field slitless spectroscopy mode. These tentative He II detections are promising for the ongoing *James Webb Space Telescope (JWST)* NIRSpec GTO program, JADES, which will observe the He II $\lambda$1640 line of galaxies in the EoR (leads, M. Rieke & P. Ferruit); and future observations with extremely-large (≥ 20 m in diameter) ground-based telescopes. Thus, in principle, the nebular He II lines can be used to identify and constrain the properties of sources of hard ionizing radiation in SF galaxies over a large range of distances. For this purpose, it is crucial to understand which are main sources of these photons in these galaxies. In sect. 1.2, we summarize some of the work attempting to constrain the properties of sources of He$^+$-ionizing radiation in nearby galaxies.

### 1.2 Sources of He$^+$-ionizing photons in nearby SF galaxies.

Two of the nearest, most metal-poor starburst galaxies known, I Zw 18 and SBS 0335-052E, show strong extended nebular He II emis-sion (e.g., Kehrig et al. 2015; Izotov et al. 2006; Kehrig et al. 2018). In these two galaxies, AGN, high-mass X-ray binaries (HMXB), and fast radiative shocks have been excluded as dominant sources responsible for the nebular He II emission (Kehrig et al. 2021; Wofford et al. 2021). Of the two galaxies, SBS 0335-052E is the most luminous in integrated nebular He II emission. Wofford et al. (2021) use observations of this galaxy with the Cosmic Origins Spectrograph (COS, PI: Wofford, PID: 13788) on the *HST*, and the models of Gutkin et al. (2016), which account for single non-rotating stars of up to 300 $M_\odot$ and the ionized gas, to show that simultaneously reproducing the fluxes of all high-ionization UV lines observed with COS, including the He II $\lambda$1640 emission line, requires an unphysically-low metallicity. Wofford et al. (2021) also present observations of the galaxy with the Multi-Unit Spectroscopic Explorer (MUSE) on the *VLT* and show that the observed He II $\lambda$4686 / H$\beta$ ratio cannot be reproduced with SSP BPASS v2.1 models that account for massive star evolution in close binaries (Eldridge et al. 2017) and the ionized gas (Xiao et al. 2018).

⋆ E-mail: jasixtos @ astro.unam.mx

Senchyna et al. (2020) argue that HMXB populations may not be sufficient to account for the observed He II line strengths in nearby metal-poor galaxies, and that revised stellar wind models or inclusion of softer X-ray sources may be needed. Stars that have been stripped of their envelopes via an interaction with a binary companion and which emit a significant portion of their radiation as ionizing photons (Götberg et al. 2019) have been proposed as alternative sources of $He^+$-ionizing photons in SF galaxies (Senchyna et al. 2017). Evidence for the existence of these stars is hard to obtain but some has been collected (Wang et al. 2021). However, comparisons of observations with predictions from populations synthesis models that account for stripped-binary products have yet to demonstrate that such products generate sufficient $He^+$-ionizing flux to explain the observations of SBS 0335-052E, for instance. As an alternative mechanism, Garnett et al. (1991) suggest that fast radiative shocks due to supernova explosions can produce relatively strong He II emission in giant H II regions under certain conditions. Plat et al. (2019) successfully reproduce high-ionization UV nebular emission lines of metal-poor galaxies, including He II lines, with a combination of fast radiative shocks and photoionization by single non-rotating massive stars. In the latter work, massive stars are needed, at least partially, but which is the stellar type that contributes the most to the $He^+$-ionizing flux?

### 1.3 Nebular He II due to Wolf-Rayet stars.

Wolf-Rayet (WR) galaxies (Schaerer et al. 1999) are SF galaxies that show signatures of the presence of WR stars. Wolf-Rayet stars are descended from O-type stars with initial masses of $\gtrsim 25\,M\odot$ (depending on metallicity) and are divided into those with strong optical lines of helium and nitrogen (WN subtype) and those with strong helium, carbon, and oxygen (WC and WO subtypes). The WN subtype is further divided based on the value of N III-N v and He I-He II optical-line ratios, into ranges from WN2 to WN5 for "early WN" (WNE) stars and WN7 to WN9 for "late WN" (WNL) stars, with WN6 stars being either early or late-type (Crowther 2007).

In order to produce detectable stellar (broad) He II emission, at least one of the following conditions must be met: the stars must be hot to doubly ionize He and the mass-loss rates must be high to produce dense winds. Since classical (He-burning) WR stars (cWR) meet these two conditions, the presence of broad He II is generally thought to indicate the presence of these stars. In addition, H-rich Very Massive Stars (VMS), which have masses of $M > 150\,{\rm M}_\odot$, have WR features as well. Wind signatures from VMS have been detected in 30 Dor (Crowther et al. 2016; Brands et al. 2022) and two starburst galaxies (Wofford et al. 2014; Smith et al. 2016). For reference, seven VMSs dominate the stellar He II $\lambda$1640 and C IV 1550$\lambda$ emission of 70 FUV-bright stars in the core of star cluster NGC 2070, located in the Large Magellanic Cloud (LMC; Crowther et al. 2016).

Some nearby SF galaxies with nebular He II $\lambda$4686 detections are accompanied by an underlying blue bump (e.g., Guseva et al. 2000; López-Sánchez & Esteban 2010; Kehrig et al. 2016; Mayya et al. 2020) that is generally associated with He II $\lambda$4686 blended with additional contributions from N III $\lambda$4640 in WN stars and/or C III/C IV 4650 in WC stars. These WR galaxies may also show a red WR bump, composed by the broad C IV $\lambda$5808 line that is mainly observed in WC stars (López-Sánchez & Esteban 2010).

When both stellar and nebular He II $\lambda$4686 emission are observed in SF galaxies, the nebular component is sometimes attributed, at least partially, to the escape of $He^+$ ionizing photons from the WR star winds (e.g. Guseva et al. 2000; Kehrig et al. 2016; Mayya et al. 2020). Schaerer & Vacca (1998) synthesized the nebular and WR He II $\lambda$4686 emission from low-metallicity young starbursts ($Z \leq Z_\odot/5$) and found that the nebular He II emission is associated with the presence of WC/WO stars and/or hot WN stars evolving to become WC/WO stars. For WR stars, the production of $He^+$-ionizing flux is a matter of wind strength. As a rule of thumb, stellar evolution models predict $He^+$-ionizing flux for WN2 (i.e., early, nitrogen-rich, classical WRs), WN3ha (i.e., early, nitrogen-rich WRs showing H-lines in emission and absorption), and WO (i.e., hot evolved stars showing strong oxygen emission lines) stars, but not for other WRs or VMS. In some works however, the nebular He II emission is associated with the presence of WNL stars (e.g. Mayya et al. 2020). Do observations of the Small Magellanic Cloud (SMC) show the same result?

### 1.4 NGC 346.

Due to its proximity, massive-star content, and low metallicity, the young massive cluster (YMC; Portegies Zwart et al. 2010) NGC 346 is a unique target to address this question.

NGC 346 is located in the SMC, at only 61±1 kpc (Hilditch et al. 2005), specifically in star-forming complex LHA 115-N66, whose radius is 3.5' (Relaño et al. 2002) and is the brightest H II region of the SMC (Henize 1956). The cluster has an age of ∼ 3 ± 1 Myr (Sabbi et al. 2007) and a present-day mass function in the range from 0.6 to 60 $M_\odot$ that is in agreement with the Salpeter stellar initial mass function (Sabbi et al. 2008). Its total mass, $3.9 \times 10^5\,M_\odot$ (Sabbi et al. 2008, Table 1), makes it the most massive YMC of the SMC (Gouliermis & Hony 2015) and comparable in mass to spatially unresolved YMCs in some nearby star-forming and starburst galaxies (e.g., Adamo et al. 2017). The ionized-gas oxygen abundance of NGC 346 is about one third that of the sun's photosphere[1] (Valerdi et al. 2019, hereafter V19). The latter value is in agreement with the oxygen abundance of massive stars in the SMC (Bouret et al. 2013) and specifically NGC 346 (Rickard et al. 2022).

---

[1] We adopt the photospheric solar oxygen abundance of Asplund et al. (2009), which is 12+log(O/H)=8.69.

### 1.5 HD 5980.

NGC 346 hosts a multiple massive-star system, HD 5980, which is of great interest for understanding massive-star evolution and binary-black hole formation. HD 5980 is composed of three massive stars. Its first component, star A, is a massive ($60 \pm 10\,M_\odot$) star that went through the eruptive luminous blue variable (LBV) phase during 1993-1994, and that now has a spectrum corresponding to that of a WN6h. Its second component, star B, is a close, eclipsing companion of similar spectral type (WN6-7) and mass ($66\pm10\,M_\odot$). The spectral types are those from Shenar et al. (2016). The masses of stars A and B are determined in Koenigsberger et al. (2014). Stars A and B orbit around each-other with a period and eccentricity of $P_{AB}$ = 19.3 days and $e$ = 0.3, respectively. The spectrum of HD 5980 has a third component, star C, believed to be itself a binary system containing a late O-type supergiant. This is shown in Hillier et al. (2019), who used CMFGEN (Hillier & Lanz 2001) models to fit *HST*/STIS UV observations of HD 5980 (PI: Koenigsberger, PID: 13373, date of observation: March 24, 2014), in particular, its broad He II $\lambda 1640$ emission and C IV $\lambda\lambda 1548, 1551$ and N IV $\lambda\lambda 3479$–3485 P-Cygni like profiles. Using the 2014 data, Hillier et al. (2019) found that when Star A (the LBV) eclipses Star B, the fitted mass-loss rate and luminosity have the lowest values ever determined for such spectra. They also found that as the mass loss rate decreases, the difference between the observed and model P-Cygni profiles increases. They suggest that the discrepancy between the model and the observation could be due to an asymmetry of the wind that is not included in the model and that is due to the perturbation of the winds of the LBV and WR star by the radiation of an O supergiant plus the wind-wind interaction.

### 1.6 Nebular He II observations in NGC 346.

NGC 346 is the target of two large massive-star spectroscopic surveys, one that is ongoing ( "Ultraviolet Legacy Library of Young Stars as Essential Standards, ULLYSES", Roman-Duval et al. 2020); and the other that recently completed ("X-Shooting ULLYSES, XShootU", Vink et al., in prep.). ULLYSES is a Director's Discretionary program that is collecting data for about 250 OB stars in low metallicity regions of Local Group galaxies, using either COS or STIS on *HST*, and far- and near-UV gratings: COS G130M, COS G160M, STIS E140M, COS G185M, and STIS E230M. XShootU obtained medium-spectral-resolution spectroscopy with the X-Shooter instrument on the *VLT*. This slit-fed (11″ slit length) spectrograph provides simultaneous coverage of the wavelength region between $300 - 2500$ nm, divided into three arms; UVB ($300 \lesssim \lambda \lesssim 500$ nm), VIS ($500 \lesssim \lambda \lesssim 1000$ nm), and NIR ($1000 \lesssim \lambda \lesssim 2500$ nm). Unfortunately, the apertures of the UV and optical observations from the above surveys are not optimized to study nebular He II in the vicinity of HD 5980. In addition, there are archival *HST*/STIS UV (PI: Oskinova, PID: 15112) and *VLT*/MUSE optical (PI: Hamann, PID: 098.D-0211) observations of NGC 346. However, these observations targeted the O-star population (Rickard et al. 2022) and thus avoided the area near HD 5980

On the other hand, Valerdi et al. (2019) determined the ionized-gas oxygen abundance and primordial helium abundance of the ionized nebula surrounding NGC 346 using long-slit optical observations at three different positions, obtained with the Focal Reducer Low Dispersion Spectrograph (FORS1) of the *VLT*. In order to measure the nebular He, they required accurate measurements of the nebular He$^{++}$ lines. Since the presence of stellar lines would impact the accuracy of these measurements, V19 discarded regions along the slits containing stars or broad He II emission, in particular, a region with broad He II emission that is located at a projected distance of $\sim$ 1.2 pc south of HD 5980. In this paper, we use the FORS1 observations to look for nebular He II emission south of HD 5980.

### 1.7 This paper.

The paper is organised as follows. In sect. 2, we present the FORS1 observations, determine the distance between HD 5980 and the nearest slit, and compare the seeing of the corresponding observation to this distance. Although the FORS1 observations are not optimized for massive stars, in sect. 3, we present massive-star spectra extracted from the FORS1 slits, compare a sub-sample of the spectra to archival *VLT* FLAMES observations, and use this comparison as a second check of our slit position determinations. Although our main interest is in the blue-grism FORS1 data, we also provide a brief discussion of the available red-grism spectra in the context of massive stars. In sect. 4, we characterise and analyse the He II emission observed with FORS1 and south of HD 5980. In sect. 5.2, we discuss whether nebular He II emission is expected south of HD 5980 based on predictions from individual star models. We also discuss these predictions in the context of SF galaxies. In addition, we analyse the effect of the SNR located south of HD 5980. Finally, in sect. 6, we provide a summary and conclusions.

## 2 OBSERVATIONS AND STANDARD DATA REDUCTION.

### 2.1 Observations.

We use *VLT* FORS1 observations obtained on September 09, 2002, as part of program 440 (PI Peimbert). The observations include a direct V-band image with a plate scale of 0.2"/pixel; and long slit (0.51″ × 410″) optical spectra covering three different locations in NGC 346. Spectra were obtained at each location with three grism / filter combinations: 300V / GG375 (3440 - 7600 Å, R∼700); 600B+12 (3560 - 5974 Å, R ∼ 1300); and 600R+14 / GG435 (5330 - 7485 Å, R∼1700). Hereafter, we will refer to the corresponding observations as the low-dispersion, blue-, and red-grism spectra, respectively. The resolution of the blue-grism is 230.5 km$^{-1}$ and is enough to distinguish between narrow and broad He II emission. The data were reduced using IRAF and a standard procedure (bias, dark and plane correction, wavelength calibration and flux calibration).

The left panel of Figure 1 shows the V-band image of NGC 346 with the slit footprints overlaid. Following V19, we use A, B, and C to identify the different slit positions. In the Figure, we also indicate the locations of HD 5980 and SNR B0057-724. For slit C, we show windows that are 5" in length and whose height has been enlarged for clarity. In § 4, we extract spectra from these windows in order to characterise the He II emission south of HD 5980. The right panel of Figure 1 shows a zoomed-in version of the area around HD 5980 covering windows 17 to 26 on slit C, and the position of star 1070 which is mentioned in § 3.

### 2.2 Determination of FORS1 slit positions relative to stars.

The positions of the slit centres that are reported in V19 are only approximate. We need to know the positions of the slits more accurately in order to determine if known massive stars are located within the slits. This is particularly important for better interpreting the properties of the He II emission south of HD 5980. In order to find the slit positions, we use a similar procedure to that of Wofford et al. (2011). The method relies on the fact that at the correct slit position, the stars in the direct image that are located within the slit should align with spectral traces of the 2d-spectral image. Thus, we compare the positions of the stars in the V-band image and the blue spectra. For each of slits A, B, and C (from top to bottom), the left panel of Figure 2 shows a portion of the V-band image that includes the relevant slit, while the right panel shows a portion of the 2d spectrum corresponding to that slit. The slit positions that are reported in V19 are shown with vertical-dashed lines, while the positions found in this work (hereafter, S22), are shown with vertical-solid lines. Table 1 gives the V19 and S22 slit centre positions. For the three slits, the differences in declinations between V19 and S22 are about 3.2". We estimate that our slit declinations are accurate within ±0.6" (3 pixels), which is about the width of the slit. No significan adjustment in RA was necessary.

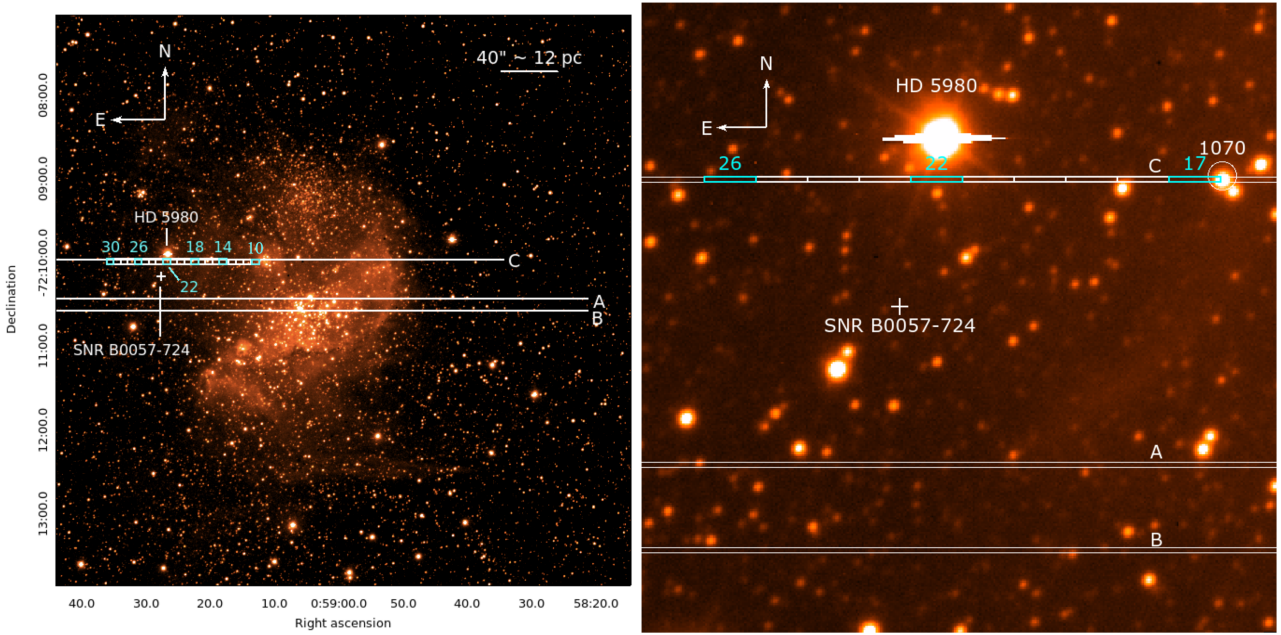

**Figure 1.** Left.–*VLT* FORS1 V-band image of NGC 346 in logarithmic scale. We overlay the footprints of the 0.51"x410" slits at the positions which were determined in this work. We indicate the positions of HD 5980 (Hillier et al. 2019) and SNR B0057-724 (Reid et al. 2006). For slit C, we show 20 spectral extraction windows of 5" in length and height enlarged for clarity, which we use to study the He II emission south of HD 5980. The cyan numbers are the window IDs. At the adopted distance of NGC 346 (61 kpc), 40" represent 11.83 pc and slit C is 1.2 pc from HD 5980. Right panel.–Zoomed-in version of the area around HD 5980 covering windows 17 to 26 on slit C and the position of star 1070, which is mentioned in § 3.

**Table 1.** Coordinates of slit centres.

| Slit | RA[a] V19 | Dec[a] V19 | RA[b] S22 | Dec[b] S22 | ΔDec[c] (V19-S22) |
|---|---|---|---|---|---|
| A | 00:59:06 | −72:10:29.3 | 0:59:06 | −72:10:25.75 | 3.5" |
| B | 00:59:06 | −72:10:37.3 | 0:59:06 | −72:10:34.1 | 3.2" |
| C | 00:59:19 | −72:10:00.7 | 0:59:19 | −72:09:57.8 | 2.9" |

[a] J2000 values from Valerdi et al. (2019).
[b] J2000 values derived in this work.
[c] Difference in declination of slit centre between V19 and S22.

In Figure 2, we use the star IDs of Dufton et al. (2019, hereafter, D19) to mark the positions of their stars within our slits. If the star is part of the ULLYSES survey, then we add an asterisk to the star ID. This is the case for HD 5980 and NGC346 435 (1001, according to D19). The figure shows that spectral traces are brighter when the star is located within the slit, and fainter when the star is either located just outside the slit (e.g., star 1064 in the top panels) or there is nebular plus stellar emission, as is the case for HD 5980 (bottom panels of Figure 2). The type of star also affects the brightness of the spectral trace. Note that the spectral trace of HD 5980 in the bottom-right panel shows strong He II $\lambda$4686 emission.

Figure 3 shows that the S22 slit positions yield a good match between intensity peaks due to the presence of stars in the direct image (middle panels) and due to the presence of spectral traces (bottom panels). The figure also shows that the V19 slit positions do not yielded good matches. The plots of Figure 3 are obtained by fixing the row number and summing the counts in the relevant columns of the image. For the top and middle panels, the columns are those within each slit. For the bottom plots, we sum the counts of columns 800 to 850, which span a wavelength interval that is free of bright emission lines (see right panels of Figure 2). Note that in Figure 3, the count sums are divided by the maximum amplitude of the peaks in each panel.

### 2.3 Seeing.

The blue-grism FORS1 observation corresponds to a Julian Date (JD) of 2452529.65. The values of the seeing during this observation were 1.14" at the beginning and 1.25" at the end of the run, i.e. smaller than 4", which is the distance from slit C to HD 5980. Thus, in principle, there should be no significant contamination from the stars in HD 5980 at the position of slit C.

## 3 MASSIVE-STAR SPECTRA.

D19 obtained stellar parameters and rotational velocities from a large sample of massive-star optical spectra obtained with the Fibre

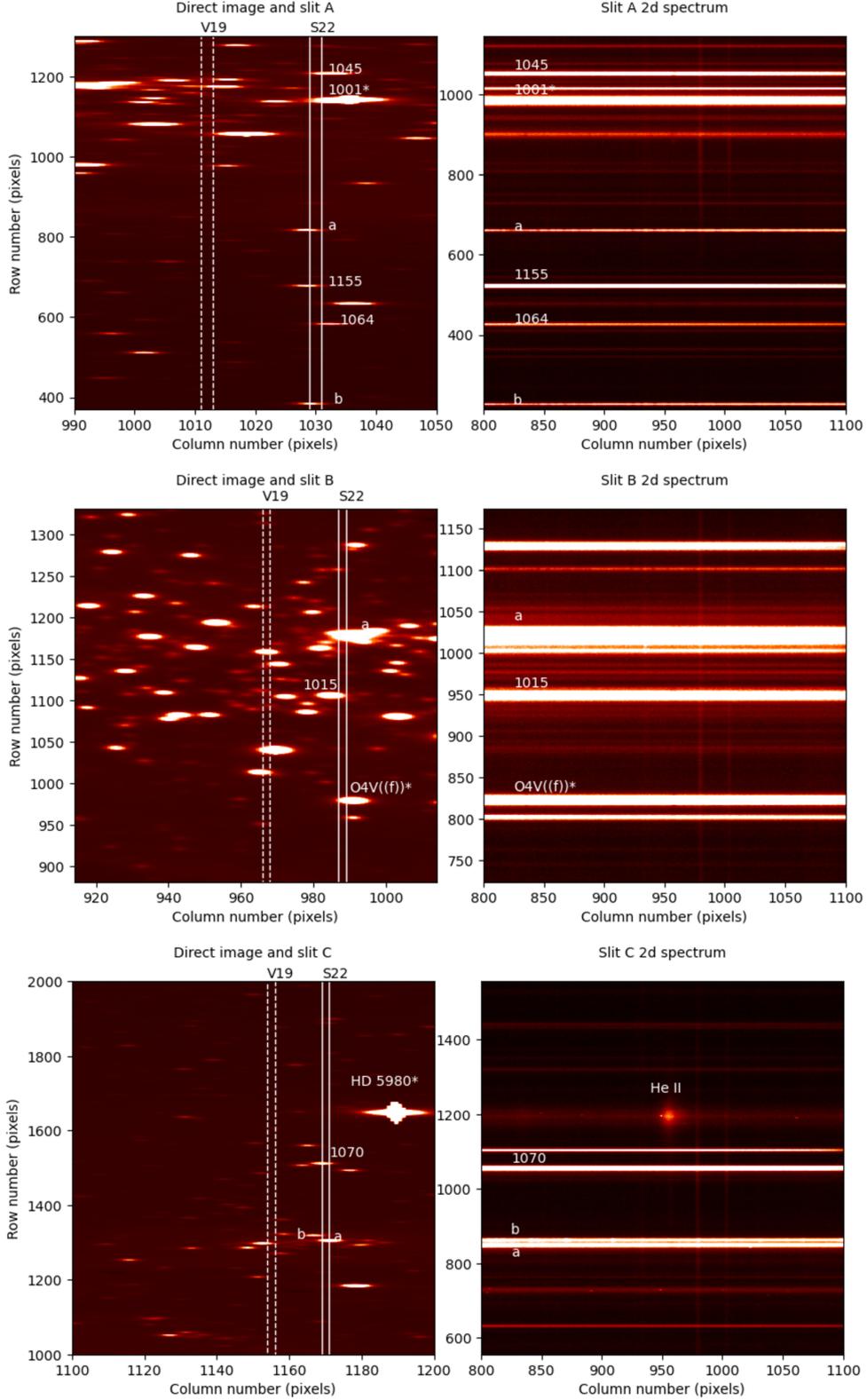

**Figure 2.** Left panels–. Portions of the direct images near slits A, B, and C from top to bottom. East is up and north is to the right. We overlay the footprints of the slits at positions V19 (dashed vertical lines) and S22 (solid vertical lines). The slit is ∼ 0.51", i.e., 2 pixels wide. Except for HD 5980 and O4 V((f)), we use the D19 IDs of known massive stars to indicate their positions. For unclassified stars with spectral traces we use lower-case letters. We add an asterisk if the known star is part of ULLYSES. Stars marked with lower-case letters are unclassified and used for reference in Figure 3. Right–. Portions of the blue grism 2d spectra showing that the S22 positions are more accurate than the V19 ones. Note the bright He II emission that is located south of HD 5980 (bottom-right panel). The direct and 2d images are stretched because the aspect ratio of the panels is one but the pixel ranges are different for the x and y axes. However, the number of rows is the same for the direct and 2d images. The number of rows shown for each slit varies in order to show relevant features.

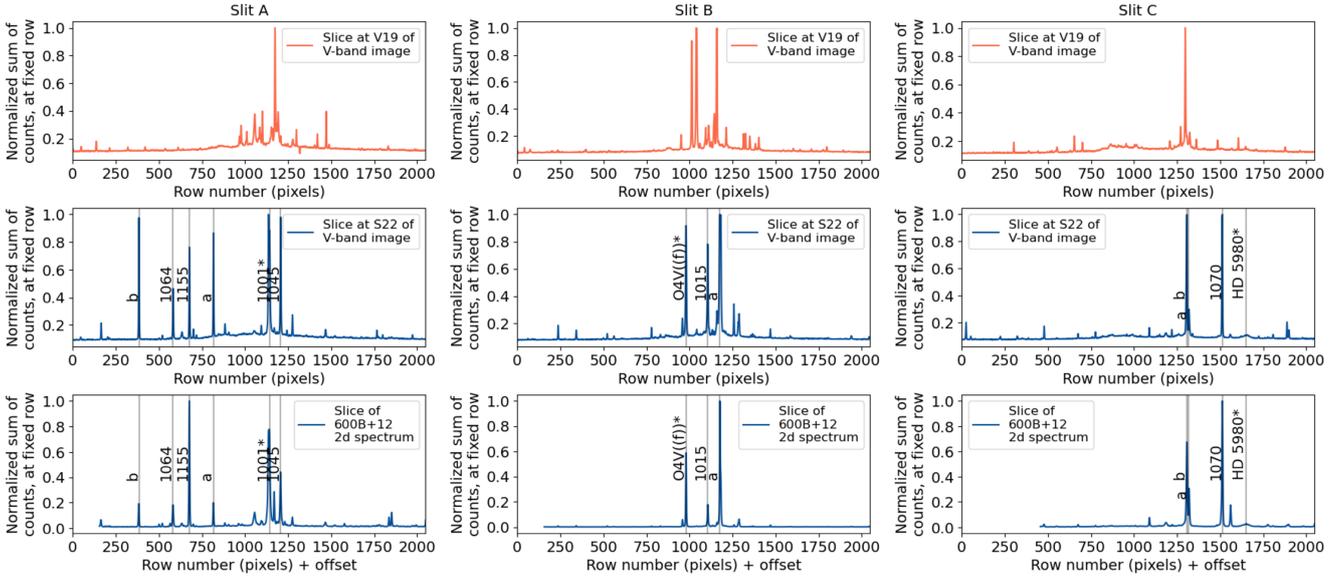

**Figure 3.** Top panels–. For every row of the V-band image, sum of the counts in the columns that span the width of the slit, when the slit is located at the V19 position. The sums are divided by the largest sum. From left to rigth we show results for slits A, B, and C. Middle panels–. Similar to the top panel but now the slit is located at the S22 position. We label the peaks corresponding to stars that are labelled in Figure 2. Bottom panels–. For every row of the 600B+12 2d spectrum, sum of the counts in columns 800 to 850. As can be seen in Figure 2, these columns are free of emission lines. The sums are divided by the largest sum. We label the peaks corresponding to stars that are labelled in Figure 2. Note that the peaks of the top and bottom panels are not aligned, contrary to the peaks of the middle and bottom panels. This is because the S22 positions are more accurate. Also note that the 2d spectrum covers bluer wavelengths than the V-band image. This affects the relative heights of the peaks in the middle and bottom panels.

Large Array Multi-Element Spectrograph (FLAMES) and the Giraffe spectrograph on the *VLT*. Some of the stars in D19 are near or within the slits used in V19. P. Dufton kindly shared these spectra with us. By comparing the latter spectra with the blue-grism FORS1 observations, we can double check the accuracy of our *VLT* FORS1 slit-position determinations.

Table 2 provides the list of stars within FORS1 slits A, B and C for which we extract spectra. The Table gives the stellar ID, spectral classification, coordinates, bottom-row pixel used in our spectral extraction, and reference for the spectral classification. The meanings of the O-star designations are given in table 1 of van der Hucht (1996). The table includes a region near multiple-star system HD 5980 and an O4 V ((f)) star which are not included in D19. We show their spectra in this section for completeness.

### 3.1 Spectral extraction.

We use a custom python routine to extract one-dimensional spectra from the two-dimensional FORS1 spectral images using boxes of 6 pixels in height. This height, which corresponds to the diameter of the *VLT* FLAMES-Medusa fibres, i.e., to 1.2", projects to 0.35 pc at the distance of the SMC. In appendix A, we show that our python extraction code yields spectra that are undistinguishable from those obtained with IRAF's apall routine.

For the O4 V ((f)) star and the stars in D19, we subtract the background. This does not include HD 5980, in which case we are interested in the nebular contribution. The box for the background spectrum is 6 pixels in height and is located in the vicinity of the spectral trace, avoiding nearby stellar sources. We use python `numpy.polyfit` to fit a 10th order polynomial to line-free portions of the sky-subtracted FORS1 spectra. We then normalize each extracted FORS1 spectrum by dividing the observed flux by the polynomial. We correct the spectra for the redshift, which we derive from the nebular [Ar IV] $\lambda$4740 emission line of NGC 346 obtained for window #14 in Figure 1. The redshift is $z = 0.00068$, corresponds to $204 \pm 4 \,\mathrm{km\,s^{-1}}$, and is within values measured from ISM components along the line of sight to the brightest stars, which go from ~ 130 km s$^{-1}$ to 210 km s$^{-1}$ (see Koenigsberber, Peimbert, et al. 2001).

Figure 4 shows the comparison between the normalised FLAMES and blue-grism FORS1 spectra. We show the FLAMES spectra at the original resolution (top blue curve) and with the same sampling as the FORS1 spectra (middle blue curve). The title of each panel gives the ID of the FORS1 slit, the star ID from D19, and the spectral and luminosity class of the star from Table 2. In the figure, we mark the rest-frame wavelengths of the strongest lines that are listed in see Table 3.

In Martins & Palacios (2017), the spectral type is determined from the ratio of the equivalent width of He I $\lambda$4471 to He II $\lambda$4542 for O4 to O9.7 stars. For O8.5 to B0 stars, the criteria from Sota et al. (2011), which include the ratio of the S III $\lambda$4552 to He II $\lambda$4542 line, are added. Figure 4 shows that for stars of these types, the FORS1 spectra have sufficient spectral resolution and signal-to-noise ratio (SNR) to estimate the ratio of the equivalent width of He I $\lambda$4471 to He II $\lambda$4542. In particular, for the Main Sequence stars (top three panels), note that the high-ionization optical He II absorption lines are absent in the B3 star and get stronger from O9 to O7 as the corresponding stars get hotter. On the other hand, the weak S III $\lambda$4552 of the O9 V star is barely detected. In addition, in Martins & Palacios (2017), the luminosity class of O stars with spectral type earlier than O8.5 is obtained from the morphology of the He II $\lambda$4686. In the case of the O4 If + O5-6 system, the quality of the FORS1 data is insufficient to appreciate the morphology of the weak line He II $\lambda$4686. In Evans et al. (2004) and D19, a combination of the equivalent width of H$\gamma$, which is clearly detected, and the B-band magnitude are used for determining the luminosity class of early B stars. Finally, the O III 5592 line has been used to determine rotational velocities of Galactic O-type stars (e.g. Simón-Díaz & Herrero 2014). This line that is not covered by the FLAMES observations is covered by the FORS1 blue-grism. However, at the low metallicity of the SMC, the line is too weak in the FORS1 data to be clearly detected. Note that the wavelength range that is shown in Figure 4 does not include the latter line. The similarities between the FLAMES and FORS1 spectra give us confidence that our FORS1 slit position determinations are sufficiently accurate for our purposes.

**Table 2.** Massive stars and regions near them for which we extract *VLT* FORS1 spectra, grouped by luminosity class.

| Slit[a] | ID[b] | Classification[c] | RA[d] | Dec[d] | y-pixel[e] | Ref[f] |
|---|---|---|---|---|---|---|
| (1) | (2) | (3) | (4) | (5) | (6) | (7) |
| A | 1164 | B3 V | 0:58:40.140 | -72:10:25.000 | 426 | D19 |
| A | 1045 | O9 V | 0:59:07.330 | -72:10:25.300 | 1051 | D19 |
| B | 1015 | O7 V | 0:59:02.904 | -72:10:34.569 | 935 | D19 |
| B | O4 V((f))* | O4 V((f)) | 0:58:57.419 | -72:10:33.268 | 822 | W00 |
| A | 1001* | O4 If + O5-6 | 0:59:04.499 | -72:10:24.766 | 984 | D19 |
| A | 1155 | B1.5 III | 0:58:44.280 | -72:10:25.800 | 521 | D19 |
| C | 1070 | B2 III | 0:59:20.600 | -72:09:58.000 | 1056 | D19 |
| C | HD 5980* | WN5-6 + WN4 + OI | 0:59:26.587 | -72:09:53.948 | 1194 | H19 |

[a]Slit where the star is located.
[b]ID of the star. The asterisk indicates that the star is a ULLYSES target. The names of the stars in ULLYSES are: NGC346-7 (O4V((f))) and NGC346 435 (1001). Note that 1001 is classified as an O4III(n)(f) in ULLYSES, https://ullyses.stsci.edu/ullyses-targets-smc.html.
[c]Classification from the reference given in column (7).
[d]Star coordinates from the reference given in column (7).
[e]Bottom row (pixel) of spectral extraction box.
[f]Reference for the stellar classification. D19: Dufton et al. (2019). W00: Walborn et al. (2000). H19: Hillier et al. (2019).

**Table 3.** Line list corresponding to Figure 4. The rest-frame wavelengths are in air.

| Ion | $\lambda_{\text{air}}$ |
|---|---|
| He I | 3964.73 |
| [Ne III] | 3967.79 |
| H7 | 3970.07 |
| He II | 4025.60 |
| He I | 4026.18 |
| $H_\delta$ | 4101.73 |
| He I | 4120.81 |
| He I | 4143.76 |
| He I | 4168.97 |
| He II | 4199.83 |
| He I | 4387.93 |
| $H_\gamma$ | 4340.46 |
| [O III] | 4363.21 |
| He I | 4471.47 |
| He II | 4541.59 |
| Si III | 4552.62 |
| He II | 4685.71 |
| He I | 4713.16 |

For completeness, we extract blue-grism FORS1 spectra of two additional sources that appear in Figures 2 and 3, are listed in Table 2, but are not in D19, i.e., an O4 V ((f)) star and a region in close proximity to HD 5980. Figure 5 shows the result. For the spectrum corresponding to the vicinity of HD 5980, we do not subtract the background, as we are interested in any nebular emission near this source. The bottom panel of Figure 5 and more detailed analysis of sect. 4 show that no nebular He II emission is detected in the FORS1 observations. Note that in the bottom panel of Figure 5, the narrow peaks marked with dotted-vertical lines are contaminating cosmic rays. In the latter panel, also note the detection of broad He II 4686 and N IV 4058 emission, which are characteristic of WN5-6 SMC stars (see figure 3 of Crowther & Hadfield 2006). The N III 4634-41 emission feature is very weak.

### 3.2 Red-grism spectra

Given that our main goal is the study of nebular He II emission in SF galaxies, our main interest is in the blue-grism FORS1 spectra. However, we also have red-grism FORS1 data. One of the most useful applications of the FORS1 red-grism spectra would be the study of variability in the broad stellar $H\alpha$ emission line of systems 1001 and HD 5980. Unfortunately, in the case of 1001, the stellar component is in absorption and in the case of HD 5980, the strong nebular component makes it difficult to extract an accurate stellar component. Figure 6 shows the spectra around $H\alpha$ of these two stars.

## 4 HE II EMISSION SOUTH OF HD 5980.

The FORS1 slit C is located at a projected distance of ∼ 1.2 pc south of HD 5980. As previously mentioned, the seeing during the observation with the blue-grism is smaller than this distance. In order to study the He II $\lambda$4686 emission south of HD 5980 (hereafter He II emission), we extract slit-C spectra within windows of 0.51"×5" in size, which at the distance of the SMC corresponds to boxes of 0.15 pc × 1.48 pc in size. The windows are shown in Figure 1. We extract a spectrum from each window using a custom python routine. We do not subtract the nebular emission in order to check if any contaminates the He II emission. For each window, we perform the following steps. 1) We match the fluxes of the FORS1 blue and red grisms by following the method that is described in appendix B. Note that this step is mostly required for step 2 below and for plotting FORS1 observations that use both grisms in the diagnostic diagram of sect. 5.3. 2) We transform from counts per pixel to flux units by multiplying the counts per pixel by $2.36\times10^{-15}$ erg s$^{-1}$ cm$^{-2}$ Å$^{-1}$. We obtained the latter factor by comparing flux-calibrated spectra from V19 with spectra that we extracted from the same regions. This is because the headers of the files containing the archival FORS1 2d spectra do not contain the information that is required to perform the flux calibration. 3) We correct for foreground extinction due to the Milky Way (MW) using $E(B - V) = 0.033$ mag (Schlafly & Finkbeiner 2011) and the MW extinction law of Fitzpatrick (1999). 4) We use the redshift measured from the nebular [Ar IV] $\lambda$4740 emission line in window #14, i.e., $z = 0.00068$, to correct each extracted spectrum for the redshift. 5) We correct for reddening due to dust in NGC 346 by using $E(B - V) = 0.08$ mag (Hennekemper et al. 2008) and the SMC Bar extinction law of Gordon et al. (2003).

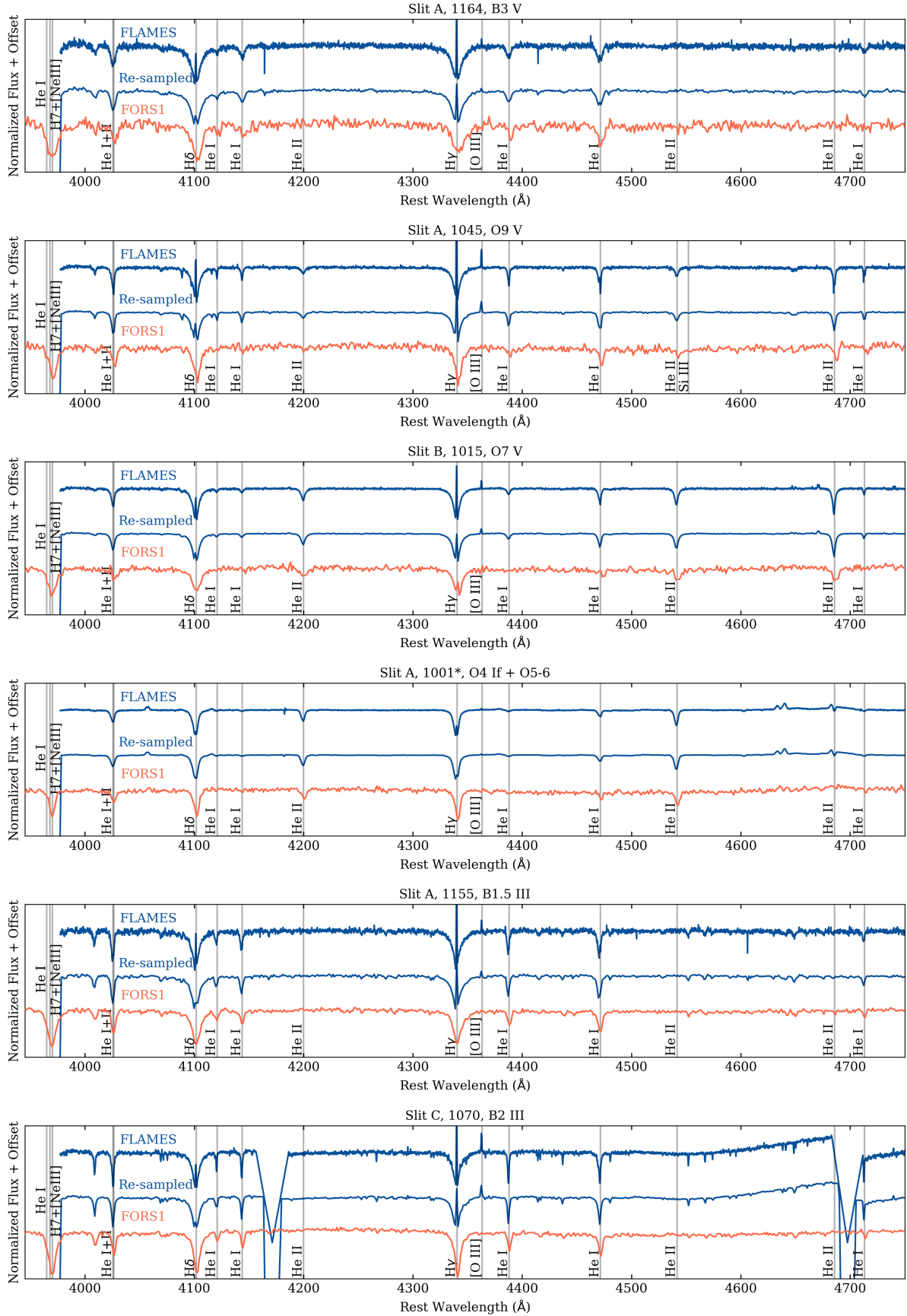

**Figure 4.** Normalised spectra from FLAMES's 1.2" fibres (top blue curves) and FORS1's 0.5"×1.2" windows (bottom red curves). We also show the FLAMES spectra re-sampled to match the FORS1 spectra (middle blue curves). We mark the rest-frame wavelengths of spectral lines listed in Table 3. In the bottom panel, the two gaps in the FLAMES spectra are due to hot pixels in the array that were removed to enable normalisation.

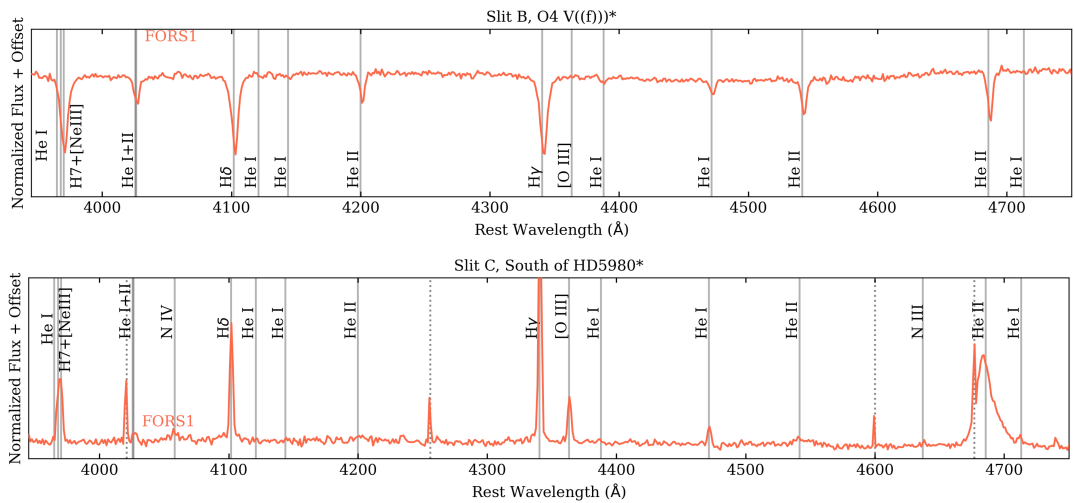

**Figure 5.** FORS1 spectra of two stars that are not in D19 but that are in ULLYSES. In the case of HD 5980, the narrow peaks marked with dotted-vertical lines are contaminating cosmic rays. Note that for HD 5980, no background subtraction was performed.

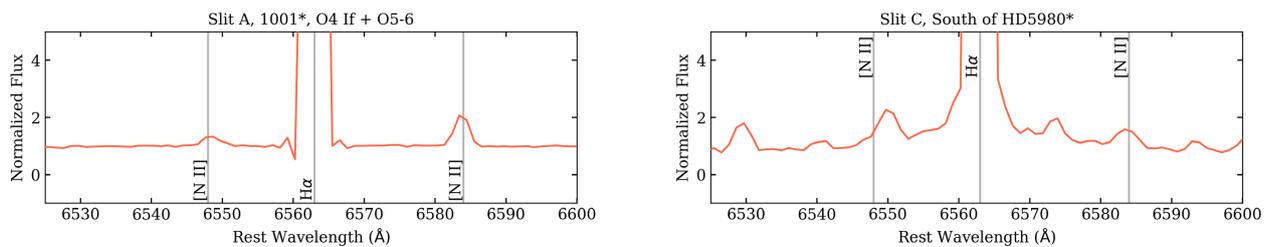

**Figure 6.** FORS1 red-grism spectra around H$\alpha$.

### 4.1 Properties of the broad He II emission lines.

For the extraction windows with a SNR ≥ 5 in the He II $\lambda$4686 line and a clear broad ($\gtrsim 1000$ km s$^{-1}$) He II component, Figure 7 shows the continuum-subtracted He II $\lambda$4686 line profiles. The fit to the continuum is a first-order polynomial that uses the wavelength range 4686 ± 100 Å and avoids the strongest emission lines. The total distance covered by the ten windows along the slit corresponds to ~ 15 pc. In Figure 7, the FWHM and velocity shift of the He II line components are given in the legend of each panel. As expected, the strongest He II emission corresponds to window 22, which is the closest to HD 5980. Two broad He II components are required to reproduce the He II profiles of windows 20 to 26, while only one broad component is necessary for the rest of the windows. In particular, the He II profile corresponding to window 22 is composed of a redshifted component with $FWHM \sim 1439$ km s$^{-1}$ and a blueshifted component with $FWHM \sim 603$ km s$^{-1}$.

### 4.2 Orbital phase during the FORS1 blue-grism observation.

In order to interpret the shape of the He II profile corresponding to window 22, we start by comparing the orbital phases corresponding to i) the eclipse between the two WN stars and ii) the date of the FORS1 blue-grism observation. The orbital period of the WR stars is 19.265 days and the date of the eclipse is JD2443158.77 (Koenigsberger et al. 2010). The latter eclipse corresponds to an orbital phase of $\phi = 0.36$ and to Star B in front of Star A, by definition of Star B. On the other hand, the date of the FORS1 observation is JD2452528, which corresponds to $\phi = 0.32$, i.e., very close to the phase of the eclipse. Hillier et al. (2019) showed that at $\phi = 0.36$, a significant amount of line emission originating in Star A is observed. This is because the wind line-emitting region has a significantly larger radius than the eclipsing disk of Star B. Therefore, even if Star B had no wind, we would still see a WR emission-line spectrum at $\phi = 0.36$ due to Star A. Not only is the date of the FORS1 observation close to the eclipse but the shape of the He II profile observed with FORS1 is qualitatively very similar to an independent STIS optical observation at $\phi = 0.36$. In the next section, we describe the *HST*/STIS observation and compare the two profiles quantitatively.

### 4.3 Comparison with STIS observation of HD 5980

The He II $\lambda$4686 emission of the multiple-star system HD 5980 was observed with *HST*/STIS and the 52"x0.2" slit, as part of program 14476 (PI: Nazé, ID: 14476, Hillier et al. 2019). The observation, which was centered on HD 5980 (unlike the FORS1 observation, which observed further south) was obtained on September 21, 2016 (JD = 2457653.06), and corresponds to the orbital phase of the eclipse ($\phi = 0.36$, Koenigsberger et al. 2010). Figure 8 shows the He II $\lambda$4686 emission line profiles corresponding to the FORS1/slit

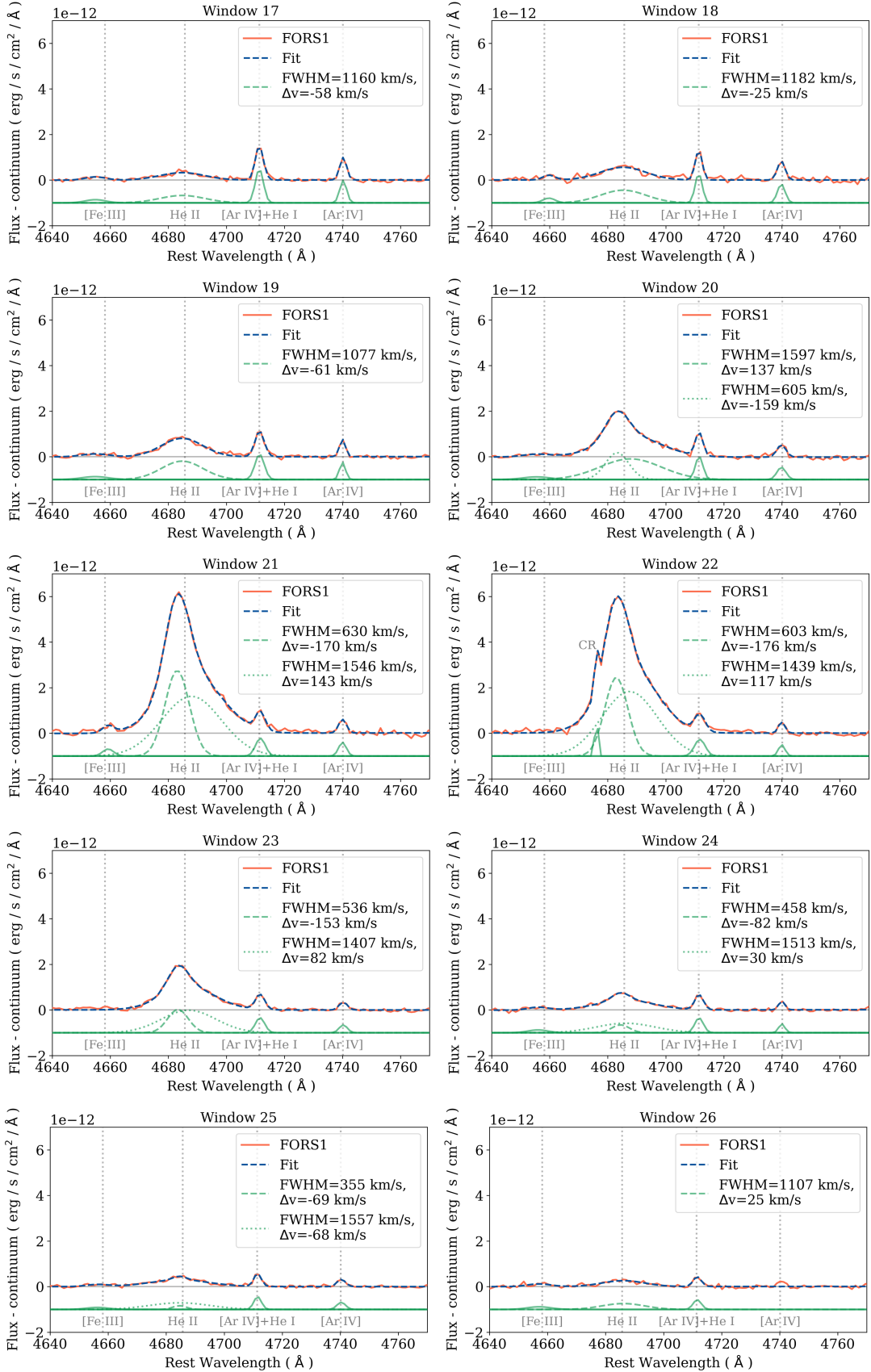

**Figure 7.** Continuum-subtracted spectra corresponding to ten windows along slit C, with a SNR ≥ 5 in the He II $\lambda$4686 line (red curves). The blue curve is a multi-gaussian fit to the observations. The green curves below the observations show the individual Gaussian components (we use a dashed curve for He II). Nearby lines are labelled underneath the green curves. The legend gives the FWHM and velocity shift of the He II line components.

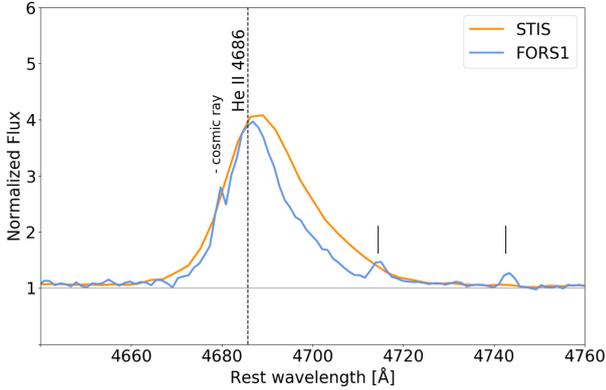

**Figure 8.** Orange curve-. HST STIS observations (aperture 53"x0.2", PI: Nazé, ID: 14476) of HD 5980, taken with G430L filter on September 21, 2016 (JD = 2457653.05913) corresponding to the eclipse (0.36, Koenigsberger et al. (2010)). Blue curve.- FORS1 observations (see 2) corresponding to a orbital phase 0.32. The black segments above the two emission lines mark the positions of the [Ar IV] + He I $\lambda$4711 and [Ar IV] $\lambda$ 4740 nebular lines in the blue spectrum, respectively.

C/window 22 and STIS observations. The figure shows that both profiles have a very similar morphology. By fitting two Gaussians to the STIS He II observation, we find a similar result than for FORS1, i.e., the best fit has a broad component with $FWHM \sim 1700 \, \mathrm{km \, s^{-1}}$ that is redshifted by $527 \, \mathrm{km \, s^{-1}}$ and a narrower component with $FWHM \sim 900 \, \mathrm{km \, s^{-1}}$ that is blueshifted by $-84 \, \mathrm{km \, s^{-1}}$.

The above result can be interpreted in two ways: i) either the seeing of the FORS1 observation is larger than the value quoted previously and with FORS1 we are looking at light coming directly from HD 5980; or ii) with FORS1 we are catching light that is being reflected by material inside or outside of slit C. We note that we are confident in the position that we determined for slit C. This is based on results shown in the bottom panels of Figures 2 and 4, as well as in the last column of Figure 3, were we show that the FORS1 and FLAMES spectra of star 1070 are very similar.

According to Shenar et al. (2016), the spectral types of the WN stars in HD 5980 are: WN6h (A) and WN6-7 (B). In Figure 9, we show spectra extracted from windows 19 to 24 (see Figure 1) that span from 4000 to 5000 Å. The curves corresponding to windows 21 and 22 show broad N IV $\lambda$4058 and He II emission, which are characteristic of the WR stars in HD 5980. This further reinforces our conclusion that we are looking at direct of reflected light from HD 5980. Also notable are the He II Pickering bumps and the broad components underlying the H I nebular lines.

## 5 IS NEBULAR HE II EMISSION EXPECTED SOUTH OF HD 5980?

Given the massive star content of HD 5980, we discuss here if nebular He II is expected around this multiple-star system.

### 5.1 Predictions from stellar evolution models

Figure 10 shows synthetic flux distributions at a distance of 10 pc from the two WN stars (top two panels) and the O I star (bottom panel) in HD 5980. The distributions are based on PoWR atmosphere model calculations (Gräfener et al. 2002; Hamann & Gräfener 2003; Sander et al. 2015) implementing the obtained parameters from the spectral analysis by Shenar et al. (2016). In each panel, we give the effective temperature ($T_{\mathrm{eff}}$) and luminosity ($L/L_\odot$) of the star, and its H I and He II ionizing rates. The figure shows that the O supergiant, despite quite some error margin due to the uncertainty in the mass-loss rate, is a far greater contributor to the He$^+$ ionizing budget of HD 5980 than the WN stars, by above three orders of magnitude. The origin of this striking difference is the much higher density in the WR winds, making them opaque to He$^+$ ionizing photons for sufficiently high mass-loss rates (e.g. Schmutz et al. 1992).

Figure 11 shows estimates for the He$^+$ ionizing rate ($Q_{\mathrm{He\,II}}$) on the Zero Age Main Sequence (ZAMS) for the SMC. To estimate the values, we calculated models from the SMC PoWR model grid (Hainich et al. 2019), but assuming the Vink et al. (2001) mass-loss rates instead of the fixed $\dot{M}$ available in the online grid. The stellar parameters are based on the SMC tracks from Brott et al. (2011). To extend the resulting relations to higher masses, an additional model for a 100 $M_\odot$ SMC ZAMS star was calculated adopting the values from corresponding BoOST SMC track (Szécsi et al. 2022). The blue lines in the figure are connecting models with the same effective temperature but different values of the surface gravity (log g) and the luminosity (log L), indicating that when going from dwarfs to supergiants, the luminosity increases, which usually increases the ionizing flux, although this trend can be reverted due to stronger mass loss consuming these photons. The orange curve is generated by using the luminosity from the SMC ZAMS inferred from Brott et al. (2011) and then interpolating the blue dataset. By adding the additional datapoint based on (Szécsi et al. 2022), we extend the curve up to initial masses of 100 $M_\odot$.

This essentially gives the estimates for log $Q_{\mathrm{He\,II}}$ as a function of L (and the initial mass). For a better reading, we indicate the values for particular initial masses with small black dots and orange labels. The absolute values of the orange curve are subject to $\dot{M}$. Here, we employed the rates of Vink et al. (2001), which are also applied in most population synthesis codes. Individual OB star analyses in the SMC often yield lower mass-loss rates (e.g., Bouret et al. 2003; Ramachandran et al. 2019; Rickard et al. 2022), while recent theoretical calculations yielded partially conflicting results (e.g. Björklund et al. 2021; Vink & Sander 2021). In case of a downward revision in mass-loss rates for the ZAMS stage, the resulting $Q_{\mathrm{He\,II}}$-values could even be higher than indicated by our orange curve in Figure 11.

Finally, the figure shows the He$^+$ ionizing rates of the three components of HD 5980. Note that the diagnostics for the O-star mass-loss rate are limited, so this star will have a higher uncertainty.

Despite all uncertainties, we can safely conclude from Figure 11 that any individual ZAMS star above 50 $M_\odot$ or any individual O supergiant hotter than $\sim$ 45 kK would outshine the whole HD 5980 system. Rickard et al. (2022) recently provided $Q_{\mathrm{He\,II}}$-values for O stars of the nearby NGC 346 cluster. Among them, only three provide a He II ionizing flux of more then $10^{43} \, s^{-1}$ and only one target – the O2 giant SSN 9 – has the capability to outshine Star C of HD 5980 with a flux of $10^{46.5} \, s^{-1}$. However, this star is in the middle of the cluster and far away from our studied region.

### 5.2 Predictions from population synthesis models.

For Simple Stellar Populations (SSPs) composed of single, non-rotating stars, and without VMS (maximum stellar mass is 100 $M_\odot$), Figure 12 shows the predicted evolution of $Q_{\mathrm{He\,II}}$ and the numbers of O, WN, and WC stars. The latter numbers are scaled down by the factors given in the legends so that all curves can be shown near to each other. The predictions are from two independent codes: Starburst99 (top panel) and Charlot & Bruzual 2019 (CB19, bottom panel).

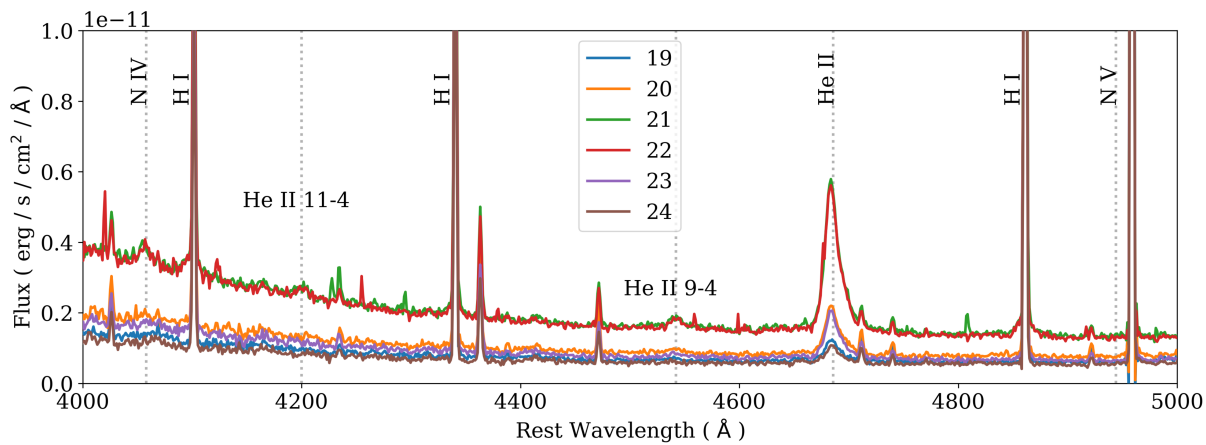

**Figure 9.** Spectra extracted from windows 19 to 24. The windows are 0.51" × 5" in size and their location along slit C is shown in Figure 1. The spectra are uncorrected for reddening due to dust but corrected for redshift. The legend gives the window number. We mark with vertical dotted lines the positions of the broad N IV $\lambda$4058 and He II $\lambda$4686 emission lines and the He II 11-4 (4200 Å) and He II 9-4 (4542 Å) Pickering bumps. Note the broad components underlying the nebular hydrogen lines.

The predictions for $Q_{\rm He\,II}$ significantly depend on the implemented evolutionary tracks, their incorporated mass-loss descriptions and the model atmospheres mapping the derived population. Hence, the absolute numbers. Starburst99 employs GENEC (Ekström et al. 2012) evolutionary tracks with $Z_\odot = 0.014$ as those for $Z = 0.008$ tracks are not yet implemented. The assigned model atmospheres, however, take the subsolar metallicity of the LMC into account and use 0.4 $Z_\odot$. The resulting output is generated as follows: i) the mass-loss rates are those inherent to the GENEC tracks for solar metallicity; ii) the star numbers are for solar-metallicity models (for O-stars, these include both main sequence and post-main-sequence stars); and iii) the number of ionizing photons is calculated from solar-metallicity tracks and atmospheres with 0.4 $Z_\odot$.

The visualizing in the top panel of Figure 12 shows that during the first 3 Myr, the value of $Q_{\rm He\,II}$ is completely determined by O-type stars (which is expected). Then, there is an increase in $Q_{\rm He\,II}$ when the number of WC's is a maximum. This is in conflict with empirical studies of WC stars, which are not contributing significant amounts of $Q_{\rm He\,II}$ due to their dense winds. However, it is a known problem in Starburst99 that the photon output for the WC does not match the observations (C. Leitherer, private communication).

The CB19 synthesis models employ PARSEC stellar evolution tracks (Bressan et al. 2012) using $Z = 0.008$ and atmospheres of roughly matching metallicity (see Plat et al. 2019, for more details). The CB19 predictions (bottom panel of Fig. 12) show an increase in $Q_{\rm He\,II}$ with the appearance of the WNLs. The WCs also contribute to this increase and at a later time the WNEs. Assuming a causality between the rising WR numbers and $Q_{\rm He\,II}$, these results are confusing as well as one would not expect the usually strong-winded WNL and WC stars to be transparent to He II ionizing photons. Both SB99 and CB99 use the PoWR atmosphere model grids (Todt et al. 2015). The mass-loss rate is one (indirect) parameter entering these grids, resulting in a set that can yield both atmospheres with and without significant He II ionizing flux.

Previous authors have found that stellar population synthesis models predict more WR stars than are observed in some nearby galaxies. In their census for hot, luminous stars in the Tarantula region in the LMC, Doran et al. (2013) found only seven classical WR stars and five very massive WNh stars residing in the core of R136. The very massive WNh stars are located in a very young region (~ 2 Myr) and will be gone by the time the cWR stars are predicted to appear by the stellar population models (~ 3 Myr, see Figure 12). The seven cWR stars observed in the wider Tarantula region (NGC 2070) are older and likely stem from the peak of star formation in this region, which was ~ 4 Myr ago according to Schneider et al. (2018). When comparing with BPASS population synthesis models, Bestenlehner et al. (2020) found that the number of observed cWR stars is actually lower than what was predicted. The Super star cluster (SSC) A in dwarf galaxy NGC 1569 has a similar age (Hunter et al. 2000) and metallicity (Kobulnicky & Skillman 1997) than the wider Tarantula region in the LMC (Mayya et al. 2020). Using Gran Telescopio Canarias MEGARA observations of SSA A Mayya et al. (2020) estimate the number of WR stars from the ratio of the observed integrated luminosity of the broad He II $\lambda$4686 line to the typical luminosity of a WNL star (~ $1.22 \times 10^{36}$ erg / s$^{-1}$ for the metallicity of the galaxy). Mayya et al. (2020) find 124 ± 11 WNL stars which is consistent with the Starburst99 predictions shown in their figure 9 (first column, middle panel) but too much compared to what has been observed in NGC 2070, even though being of similar mass. The estimates from line luminosities are very rough and the attribution to specifically WNL stars is somewhat arbitrary. Moreover, WNL stars are not expected to yield measurable amounts of He$^+$ ionizing flux (Crowther & Hadfield 2006; Sander 2022), which is at odds with the strong flux measured by Mayya et al. (2020), as there, similar to what we find here, the population models seem to predict ionizing fluxes from the WNL stage.

Given the findings in the Tarantula region and the uncertainties of the measurement by Mayya et al. (2020), it appears that the numbers of WR stars are actually overpredicted by stellar population synthesis models. However, the WR classification is formally a purely spectroscopic one, while the assignment from the population models is usually based on surface composition and temperature. Hence, the predicted stars might not actually appear as WR stars, making them much harder to detect. The suspicious coincidence between the increase of the He$^+$ ionizing flux and the number of cWR stars in fact indicates that the population synthesis models like to select weaker-winded atmosphere models which have more transparent winds with few or no emission-line signatures. In the super star cluster A region of NGC 1569, this could potentially resolve the discrepancy, but the absence of He$^+$ ionizing flux in the Tarantula region makes such an explanation invalid there. Assuming that the too low number of WR star is an effect of overestimating the self-stripping of massive star (in single star evolution), one would expect to see more RSGs instead in the Tarantula region. As evident from the compilation by Bestenlehner et al. (2020), this does not seem to be the case. Finally, the lifetimes of WRs could be shorter than assumed in the models, but given the relatively constant duration of the central He burning time in massive stars (∼ 300 kyr) this would imply that the stars spend a considerable fraction of that in a stage where they do not become visible as cWRs. However, studying this is beyond the scope of the present paper.

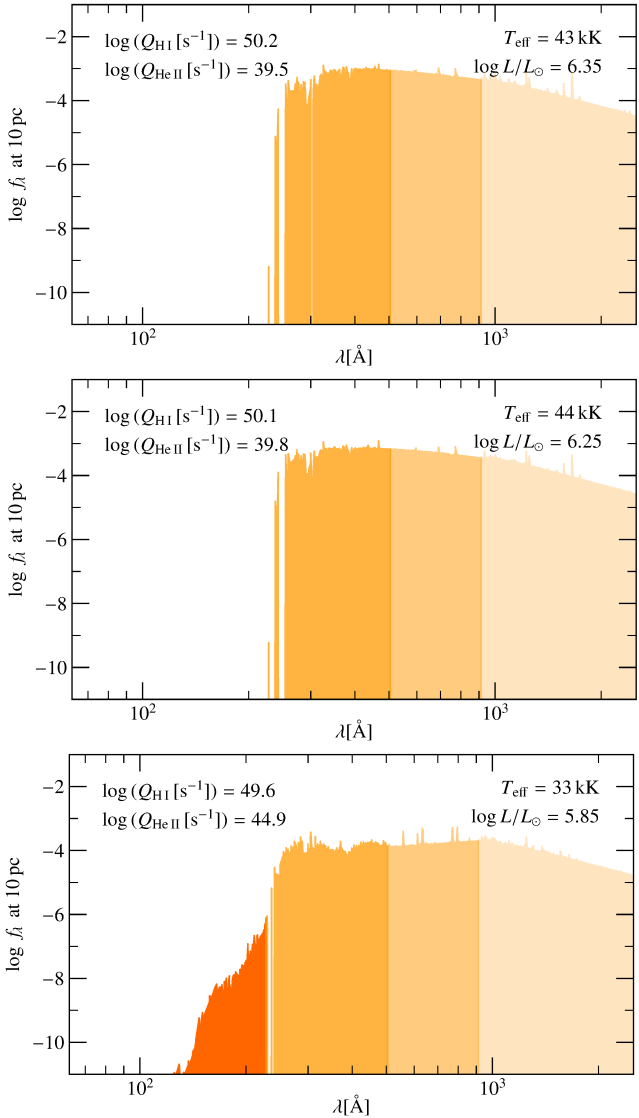

**Figure 10.** Synthetic flux distributions at a distance of 10 pc from the two WN stars (top two panesl) and the O supergiant (bottom panel) in HD 5980.

While a more detailed investigation would be necessary to precisely identify the origins of the mismatch between the empirical results and the population synthesis predictions, it seems likely that the currently implemented stellar evolution in both SB99 and CB99 yield significant populations of stars that do not have mass-loss rates that are high enough to be opaque to He II ionizing photons. Instead, the predicted stars are sufficiently hot and hydrogen-depleted to formally fall into the range where the WR grids are applied. Whether such stars would actually be classified as WR from their spectral appearance (Shenar et al. 2020), cannot be determined in the scope of this paper. Nonetheless, we can conclude that the current generation of population synthesis models seem to have significant problems to get a realistic prediction of $Q_{\mathrm{H\,II}}$, at least at subsolar metallicities. Updating $Q_{\mathrm{He\,II}}$ predictions in widely used population synthesis codes using observations like the ones presented in this work will be crucial to enable a more realistic interpretation of SF galaxies.

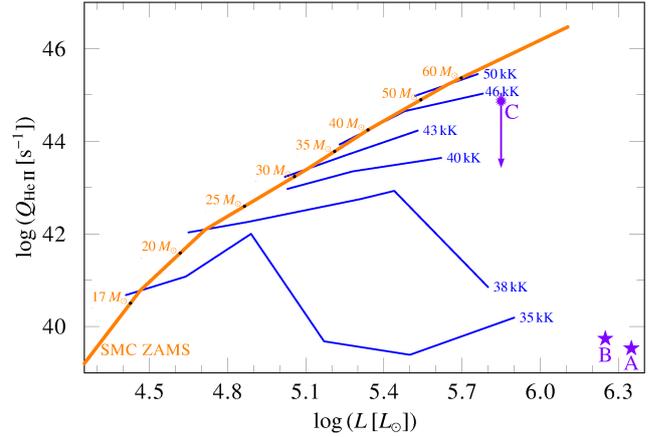

**Figure 11.** Estimates for the He$^+$ ionizing rate ($Q_{\mathrm{He\,II}}$) on the Zero Age Main Sequence (ZAMS). The blue lines connect models with the same $T_{\mathrm{eff}}$ but different values of log g and log L. The orange curve is generated by using the luminosity from the SMC ZAMS in Brott et al. (2011) and then interpolating the blue dataset (see text for more details). The positions of the stars in HD 5980 are indicated with the letters A, B, and C. We use different symbols for the two WNs and the O star.

### 5.3 What is the effect of the SNR south of HD 5980?

As mentioned in the introduction, Garnett et al. 1991 suggest that fast radiative shocks due to supernova explosions can produce relatively strong He II emission in giant H II regions under certain conditions. The extended X-ray emission around HD 5980 coincides with non-thermal radio emission, suggesting that a SNR is present in the region (Reid et al. 2006). In Figure 1, we indicate the position of SNR B0057-724, which is reported in Reid et al. (2006). The diffuse X-ray emission of the SNR reaches the region covered by our slit C (see figure 10 of Reid et al. 2006). Thus, we looked at the positions of the observations corresponding to slit C and windows 20-26 (which have broad He II) in optical diagnostic diagrams relative to the radiative shock models of Alarie & Morisset (2019). These shock models agree with predictions presented in Allen et al. (2008) at SMC metallicity but include models of lower metallicities. Figure 13 shows the diagnostic diagrams of Baldwin et al. (1981) and Veilleux & Osterbrock (1987), left and right panels, respectively. These diagrams use nebular emission line fluxes corrected for the stellar component. The red square is the integrated value of the 7 windows. For comparison, the filled circles show the SMC shocks models, presented in Alarie & Morisset (2019).

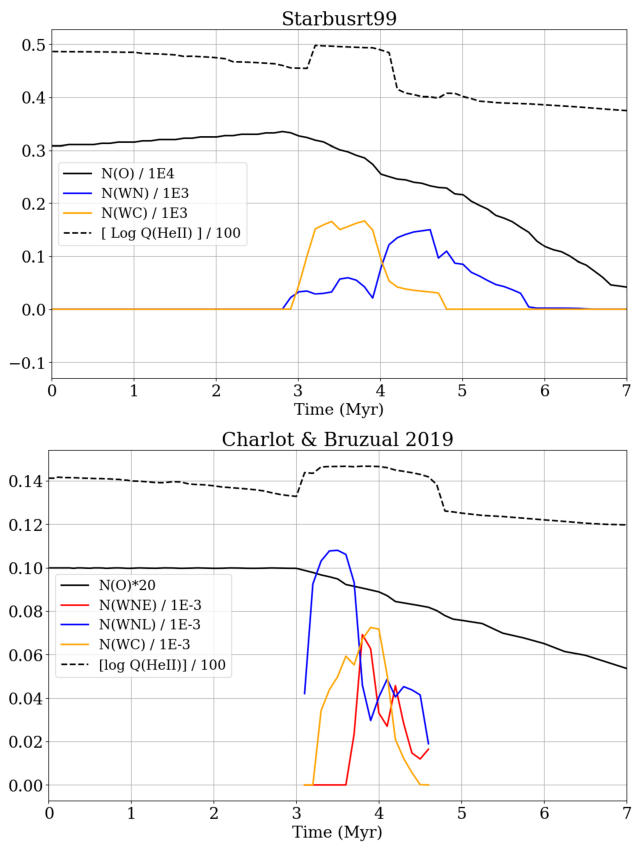

**Figure 12.** Predicted evolution of $Q_{HeII}$ and the numbers of O, WN, and WC stars for SSPs without VMSs and composed of single, non-rotating stars. The latter numbers are scaled down by the factors given in the legends so that all curves can be shown near to each other. We show predictions from two independent codes: Starburst99 (top panel, $10^6 M_\odot$ SSP) and Charlot & Bruzual 2019 (bottom panel, SSP with stellar Initial Mass Function normalized to 1 $M_\odot$). The metallicity of the stellar atmospheres approach that of the SMC (see text for more details).

The colour bar corresponds to the shock speed, taking into account values between 100 and 1000 km s$^{-1}$. We find that the optical line ratios of the spectra extracted from the windows are located far from the shock-model predictions. Thus, no evidence of the presence of the SNR is seen in these diagnostic diagrams. A caviat to this result is that the shock models do not account for any chemical enrichment from the products of the star that exploded.

The SNR is almost centered on HD 5980 and has dimensions $\sim$ 4' according to Ye et al. (1991), see his Figure 6. The X-ray emission extends South to at least 1'. See figure 1 of Nazé et al. (2002). Our slit is located 4 arcsec from HD 5980, hence, any light from the SNR should enter the slit. Velázquez et al. (2003) proposed that the observed X-rays could result from the collision of HD 5980's wind with the foreground SNR. If that were the case, then: a) there would be significant collisional excitation and ionization; and b) the SNR would be He-enriched. Thus, one could argue that there should be nebular He II emission at the position of slit C. In order to verify that we have the ability to detect nebular emission within our broad He II emission, we simulate the presence of nebular He II emission by adding the flux of the nearby, nebular [Ar IV] $\lambda 4740$ Å to the flux of He II $\lambda 4686$ Å line. Hereafter, we refer to the resulting profile as the combined profile. We chose this [Ar IV] line based on figure 1 of López-Sánchez & Esteban (2010), where the nebular He II lines have fluxes that are similar to those of the [Ar IV] line. We fit the combined He II profile, finding that a third emission component is now necessary in order to achieve a good fit. This is shown in the insets of Figure 14, where the top inset corresponds to the fit without the third component and the bottom panel to the fit with it. We conclude that we have the ability to detect nebular He II emission with the strength of the [Ar IV] $\lambda 4740$ Å line. However, we do not observe such emission.

## 6 SUMMARY AND CONCLUSIONS

We use archival long-slit observations obtained with FORS1 on the *VLT* to look for nebular He II $\lambda 4686$ emission south of the WN6-7 + WN6h close binary in HD 5980. We only find broad He II $\lambda 4686$ emission, as far as $\sim$ 7.6 pc from the binary. A comparison with observations obtained with STIS on the *HST*, at a similar orbital phase, shows that the FORS1 broad He II emission is likely contamination from multiple-star system HD 5980. We use models to show that no significant He$^+$-ionizing flux is expected from the WN stars in HD 5980 and that when similar stars are present in a coeval stellar population, the O stars are far greater emitters of He$^+$-ionizing radiation.

We also compared the spectra of known massive stars within the FORS1 slits with published *VLT* FLAMES spectra of higher quality. This comparison gives us confidence in the revised FORS1 slit positions relative to V19, which we provide in this paper. The FORS1 red-grism observations of these stars did not add anything new to what is already known about them.

With regards to the SNR near HD 5980, we find that it is unlikely that it significantly contributes to nebular He II emission at the position of FORS1 slit C. We detect none at the level of the flux of the nearby [Ar IV] $\lambda 4740$ Å.

**ACKNOWLEDGEMENTS**

We thank the referee for comments that really helped improve the quality of this paper. AW and AS acknowledge the support of UNAM via grant agreement PAPIIT no. IA102120 and IN106922. AACS is funded by the Deutsche Forschungsgemeinschaft (DFG, German Research Foundation) in the form of an Emmy Noether Research Group - Project-ID 445674056 (SA4064/1-1, PI Sander) and acknowledges additional support from a DFG collaborative research center – Project-ID 138713538 – SFB 881 ("The Milky Way System", subproject P04). We thank: Gloria Koenigsberger, John Hillier, and Yael Nazé for useful discussions that led to the current interpretation of the FORS1 He II profile south of HD 5980; Philip Dufton, Danny Lennon, and Chris Evans for useful discussions about the known massive star content in the region covered with FORS1, and for providing the normalised *VLT* FLAMES spectra of the stars in common with this work; Claus Leither, Gustavo Bruzual, and Stéphane Charlot, for their help with the output of their population synthesis models; Jay Ghallagher, Lidia Oskinova, and Matthew Rickard for helping establish that the available *HST* STIS and *VLT* MUSE observations of NGC 346 do not include the region around HD 5980; and finally, Artemio Herrero, Miriam García, and Norberto Castro for very helpful information that enriched the discussion of Figures 4 and 5.

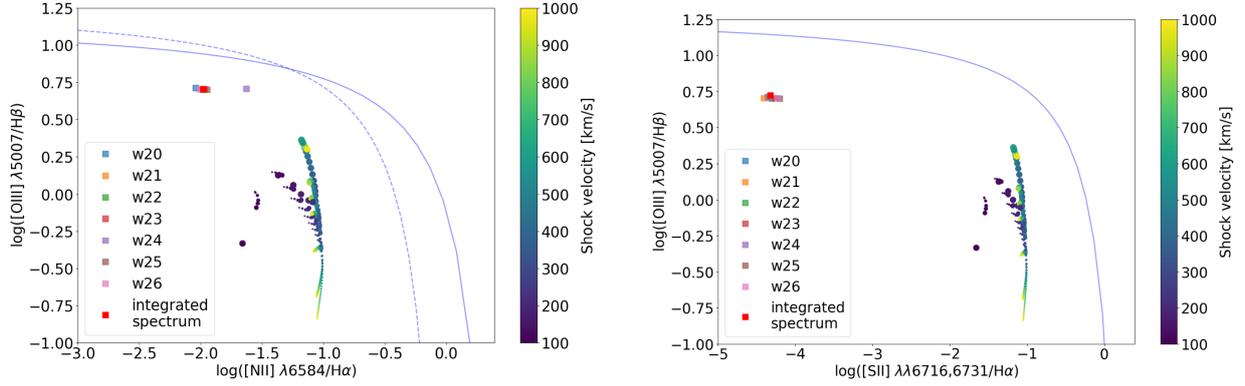

**Figure 13.** Left–. [O III] $\lambda$5007 / H$\beta$ versus [N II] $\lambda$6584 / H$\alpha$ diagnostic diagram (Baldwin et al. 1981). The dashed and solid blue lines represent the Kauffmann et al. (2003) and Kewley et al. (2001) curves for separating star forming objects from AGN, respectively. Right–. [O III] $\lambda$5007 / H$\beta$ versus [S II] $\lambda\lambda$6717, 6731/H$\alpha$ diagnostic diagram (Veilleux & Osterbrock 1987). The solid blue line is the Kewley et al. (2001) curve for separating star forming objects from AGN. In both panels, the squares represent values measured in windows of slit C with two gaussian components (Figure 7) and SNR$\geq$ 5 in the He II $\lambda$4686 line (Figure 1), and the red-filled square is the value measured after integrating the spectra of the 7 windows. We also show the shock models of (Alarie & Morisset 2019) with SMC metallicity, a pre-shock density of $n_0 = 1$ cm$^{-3}$, shock speeds between 200 and 1000 km s$^{-1}$ (see colour bar) and magnetic field strengths between 0.0001 and 10 $\mu$G (given by the size of the filled circle, where the size increases with the strength).

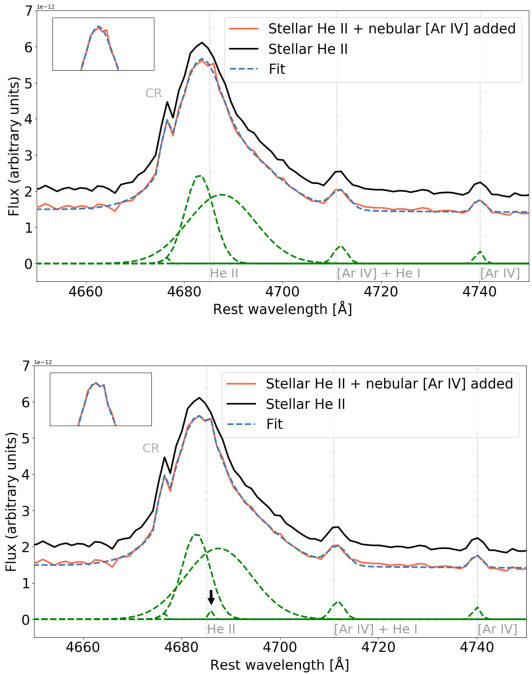

**Figure 14.** Simulation of stellar plus nebular He II profile (red curves in both panels). For simulating a nebular component, we add to the observed He II profile the [Ar IV] $\lambda$4740 Å line at 4686 Å (see text for more details). Two different fits to the result are presented (blue-dashed curves) corresponding to the Gaussian components shown with green-dashed curves. The black curve is the FORS1 observation, which is plotted offset for clarity. The insets in both panels show the peak of the combined emission (He II + [Ar IV]) and that the fit in the bottom panel is the best.

**DATA AVAILABILITY**

The data underlying this article will be shared on reasonable request to the corresponding author.

**REFERENCES**

Adamo A., et al., 2017, ApJ, 841, 131
Alarie A., Morisset C., 2019, Rev. Mex. Astron. Astrofis., 55, 377
Allen M. G., Groves B. A., Dopita M. A., Sutherland R. S., Kewley L. J., 2008, ApJS, 178, 20
Asplund M., Grevesse N., Sauval A. J., Scott P., 2009, ARA&A, 47, 481
Baldwin J. A., Phillips M. M., Terlevich R., 1981, PASP, 93, 5
Bestenlehner J. M., et al., 2020, MNRAS, 499, 1918
Björklund R., Sundqvist J. O., Puls J., Najarro F., 2021, A&A, 648, A36
Bouret J. C., Lanz T., Hillier D. J., Heap S. R., Hubeny I., Lennon D. J., Smith L. J., Evans C. J., 2003, ApJ, 595, 1182
Bouret J. C., Lanz T., Martins F., Marcolino W. L. F., Hillier D. J., Depagne E., Hubeny I., 2013, A&A, 555, A1
Brands S. A., et al., 2022, A&A, 663, A36
Bressan A., Marigo P., Girardi L., Salasnich B., Dal Cero C., Rubele S., Nanni A., 2012, MNRAS, 427, 127
Brott I., et al., 2011, A&A, 530, A115
Crowther P. A., 2007, ARA&A, 45, 177
Crowther P. A., Hadfield L. J., 2006, A&A, 449, 711
Crowther P. A., et al., 2016, MNRAS, 458, 624
Doran E. I., et al., 2013, A&A, 558, A134
Dufton P. L., Evans C. J., Hunter I., Lennon D. J., Schneider F. R. N., 2019, A&A, 626, A50
Ekström S., et al., 2012, A&A, 537, A146
Eldridge J. J., Stanway E. R., Xiao L., McClelland L. A. S., Taylor G., Ng M., Greis S. M. L., Bray J. C., 2017, Publ. Astron. Soc. Australia, 34, e058
Evans C. J., Howarth I. D., Irwin M. J., Burnley A. W., Harries T. J., 2004, MNRAS, 353, 601
Fitzpatrick E. L., 1999, PASP, 111, 63
Garnett D. R., Kennicutt Robert C. J., Chu Y.-H., Skillman E. D., 1991, ApJ, 373, 458
Gordon K. D., Clayton G. C., Misselt K. A., Landolt A. U., Wolff M. J., 2003, ApJ, 594, 279
Götberg Y., de Mink S. E., Groh J. H., Leitherer C., Norman C., 2019, A&A, 629, A134
Gouliermis D. A., Hony S., 2015, arXiv e-prints, p. arXiv:1511.00462
Gräfener G., Koesterke L., Hamann W. R., 2002, A&A, 387, 244
Guseva N. G., Izotov Y. I., Thuan T. X., 2000, ApJ, 531, 776
Gutkin J., Charlot S., Bruzual G., 2016, MNRAS, 462, 1757
Hainich R., Ramachandran V., Shenar T., Sander A. A. C., Todt H., Gruner D., Oskinova L. M., Hamann W. R., 2019, A&A, 621, A85

Hamann W. R., Gräfener G., 2003, A&A, 410, 993
Henize K. G., 1956, ApJS, 2, 315
Hennekemper E., Gouliermis D. A., Henning T., Brandner W., Dolphin A. E., 2008, ApJ, 672, 914
Hilditch R. W., Howarth I. D., Harries T. J., 2005, MNRAS, 357, 304
Hillier D. J., Lanz T., 2001, in Ferland G., Savin D. W., eds, Astronomical Society of the Pacific Conference Series Vol. 247, Spectroscopic Challenges of Photoionized Plasmas. p. 343
Hillier D. J., Koenigsberger G., Nazé Y., Morrell N., Barbá R. H., Gamen R., 2019, MNRAS, 486, 725
Hunter D. A., O'Connell R. W., Gallagher J. S., Smecker-Hane T. A., 2000, AJ, 120, 2383
Izotov Y. I., Schaerer D., Blecha A., Royer F., Guseva N. G., North P., 2006, A&A, 459, 71
James B. L., et al., 2022, arXiv e-prints, p. arXiv:2206.01224
Kauffmann G., et al., 2003, MNRAS, 346, 1055
Kehrig C., Vílchez J. M., Pérez-Montero E., Iglesias-Páramo J., Brinchmann J., Kunth D., Durret F., Bayo F. M., 2015, ApJ, 801, L28
Kehrig C., et al., 2016, MNRAS, 459, 2992
Kehrig C., Vílchez J. M., Guerrero M. A., Iglesias-Páramo J., Hunt L. K., Duarte-Puertas S., Ramos-Larios G., 2018, MNRAS, 480, 1081
Kehrig C., Guerrero M. A., Vílchez J. M., Ramos-Larios G., 2021, ApJ, 908, L54
Kewley L. J., Dopita M. A., Sutherland R. S., Heisler C. A., Trevena J., 2001, ApJ, 556, 121
Kobulnicky H. A., Skillman E. D., 1997, ApJ, 489, 636
Koenigsberger G., Georgiev L., Hillier D. J., Morrell N., Barbá R., Gamen R., 2010, AJ, 139, 2600
Koenigsberger G., Morrell N., Hillier D. J., Gamen R., Schneider F. R. N., González-Jiménez N., Langer N., Barbá R., 2014, AJ, 148, 62
López-Sánchez Á. R., Esteban C., 2010, A&A, 516, A104
Martins F., Palacios A., 2017, A&A, 598, A56
Mayya Y. D., et al., 2020, MNRAS, 498, 1496
Nazé Y., Hartwell J. M., Stevens I. R., Corcoran M. F., Chu Y. H., Koenigsberger G., Moffat A. F. J., Niemela V. S., 2002, ApJ, 580, 225
Plat A., Charlot S., Bruzual G., Feltre A., Vidal-García A., Morisset C., Chevallard J., Todt H., 2019, MNRAS, 490, 978
Portegies Zwart S. F., McMillan S. L. W., Gieles M., 2010, ARA&A, 48, 431
Ramachandran V., et al., 2019, A&A, 625, A104
Reid W. A., Payne J. L., Filipović M. D., Danforth C. W., Jones P. A., White G. L., Staveley-Smith L., 2006, MNRAS, 367, 1379
Relaño M., Peimbert M., Beckman J., 2002, ApJ, 564, 704
Rickard M. J., et al., 2022, arXiv e-prints, p. arXiv:2207.09333
Roman-Duval J., et al., 2020, Research Notes of the American Astronomical Society, 4, 205
Sabbi E., et al., 2007, AJ, 133, 44
Sabbi E., et al., 2008, AJ, 135, 173
Sander A. A. C., 2022, arXiv e-prints, p. arXiv:2211.05424
Sander A., Shenar T., Hainich R., Gímenez-García A., Todt H., Hamann W. R., 2015, A&A, 577, A13
Saxena A., et al., 2020, A&A, 636, A47
Schaerer D., Vacca W. D., 1998, ApJ, 497, 618
Schaerer D., Contini T., Pindao M., 1999, A&AS, 136, 35
Schlafly E. F., Finkbeiner D. P., 2011, ApJ, 737, 103
Schmutz W., Leitherer C., Gruenwald R., 1992, PASP, 104, 1164
Schneider F. R. N., et al., 2018, Science, 359, 69
Senchyna P., et al., 2017, MNRAS, 472, 2608
Senchyna P., Stark D. P., Mirocha J., Reines A. E., Charlot S., Jones T., Mulchaey J. S., 2020, MNRAS, 494, 941
Shenar T., et al., 2016, A&A, 591, A22
Shenar T., Gilkis A., Vink J. S., Sana H., Sander A. A. C., 2020, A&A, 634, A79
Shirazi M., Brinchmann J., 2012, MNRAS, 421, 1043
Simón-Díaz S., Herrero A., 2014, A&A, 562, A135
Smith L. J., Crowther P. A., Calzetti D., Sidoli F., 2016, ApJ, 823, 38
Sobral D., et al., 2019, MNRAS, 482, 2422
Sota A., Maíz Apellániz J., Walborn N. R., Alfaro E. J., Barbá R. H., Morrell N. I., Gamen R. C., Arias J. I., 2011, ApJS, 193, 24
Stark D. P., et al., 2015, MNRAS, 454, 1393
Szécsi D., Agrawal P., Wünsch R., Langer N., 2022, A&A, 658, A125
Todt H., Sander A., Hainich R., Hamann W. R., Quade M., Shenar T., 2015, A&A, 579, A75
Valerdi M., Peimbert A., Peimbert M., Sixtos A., 2019, ApJ, 876, 98
Veilleux S., Osterbrock D. E., 1987, ApJS, 63, 295
Velázquez P. F., Koenigsberger G., Raga A. C., 2003, in Reyes-Ruiz M., Vázquez-Semadeni E., eds, Revista Mexicana de Astronomia y Astrofisica Conference Series Vol. 18, Revista Mexicana de Astronomia y Astrofisica Conference Series. p. 150
Vink J. S., Sander A. A. C., 2021, MNRAS, 504, 2051
Vink J. S., de Koter A., Lamers H. J. G. L. M., 2001, A&A, 369, 574
Walborn N. R., Lennon D. J., Heap S. R., Lindler D. J., Smith L. J., Evans C. J., Parker J. W., 2000, PASP, 112, 1243
Wang L., Gies D. R., Peters G. J., Götberg Y., Chojnowski S. D., Lester K. V., Howell S. B., 2021, AJ, 161, 248
Wofford A., Leitherer C., Chandar R., 2011, ApJ, 727, 100
Wofford A., Leitherer C., Chandar R., Bouret J.-C., 2014, ApJ, 781, 122
Wofford A., Vidal-García A., Feltre A., Chevallard J., Charlot S., Stark D. P., Herenz E. C., Hayes M., 2021, MNRAS, 500, 2908
Xiao L., Stanway E. R., Eldridge J. J., 2018, MNRAS, 477, 904
Ye T., Turtle A. J., Kennicutt R. C. J., 1991, MNRAS, 249, 722
van der Hucht K. A., 1996, in Vreux J. M., Detal A., Fraipont-Caro D., Gosset E., Rauw G., eds, Liege International Astrophysical Colloquia Vol. 33, Liege International Astrophysical Colloquia. p. 1

## APPENDIX A: SPECTRAL EXTRACTION ROUTINE.

In Figure A1, we show spectra corresponding to slit C and window 22 of Figure 1, which were extracted with our custom Python routine (thick coloured curves) and IRAF's apall routine (thin grey curves). Since the spectra are undistinguishable from each other, we can confidently use our custum python routine for the spectral extractions used in this work.

## APPENDIX B: MATCHING THE FLUX LEVELS OF THE BLUE AND RED GRISMS.

As shown in the top-panel of Figure B1, which corresponds to the spectra extracted from slit C at window 22 of Figure 1, the continuum level of the red spectrum (red curve) does not match the continuum level of the blue spectrum (blue curve). The top panel of Figure B1 is just an example but this is true in general. Thus, a correction needs to be applied. For this purpose, we use the low dispersion spectrum, which is shown in grey in Figure B1. We proceed as follows. First, we use Gaussian fitting and a custom python routine to obtain the fluxes, $F$, of the hydrogen H$\alpha$ and H$\beta$ emission lines in the high ($h$) and low ($l$) dispersion spectra, i.e., $F_h(H\alpha)$ and $F_h(H\beta)$, and $F_l(H\alpha)$ and $F_l(H\beta)$, respectively. If for the blue spectrum, $F_h(H\beta) > F_l(H\beta)$, then we divide the flux array of the blue spectrum by $F_h(H\beta) / F_l(H\beta)$. Otherwise, we multiply it by $F_l(H\beta) / F_h(H\beta)$. We proceed similarly for the red spectrum, but this time we use the ratio of the H$\alpha$ fluxes instead. The bottom panel of Figure B1 shows that this procedure fixes the problem.

This paper has been typeset from a TeX/LaTeX file prepared by the author.

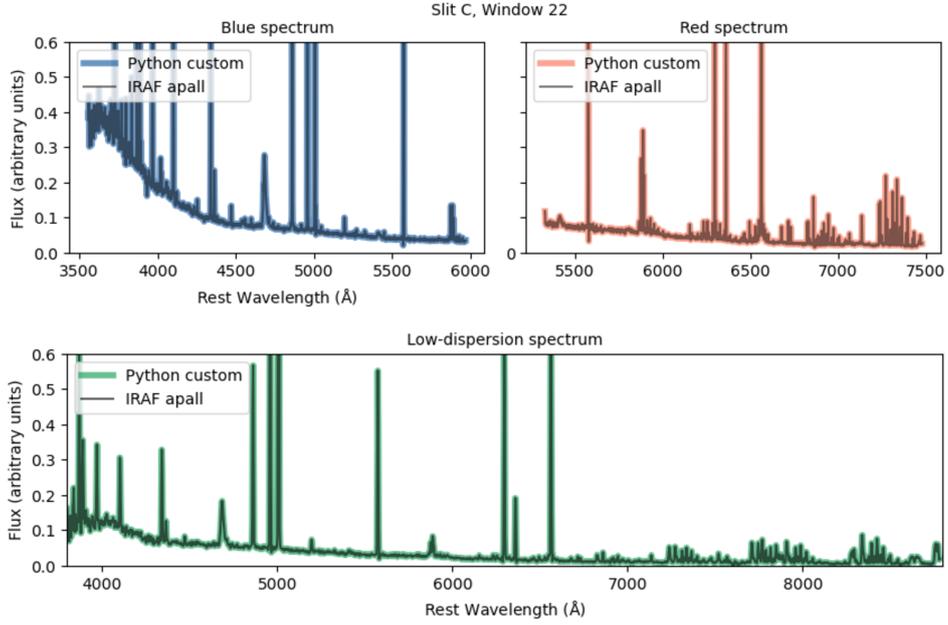

**Figure A1.** Comparison of spectra extracted from slit C and window 22 with our custom Python routine (thick coloured curves) and IRAF's apall routine (thin grey curves). The top-two panels show the blue- and red-grism spectra, respectively, while the bottom panel shows the low-dispersion grism spectrum. The figure shows that the spectra extracted with Python and IRAF are undistinguishable from each other.

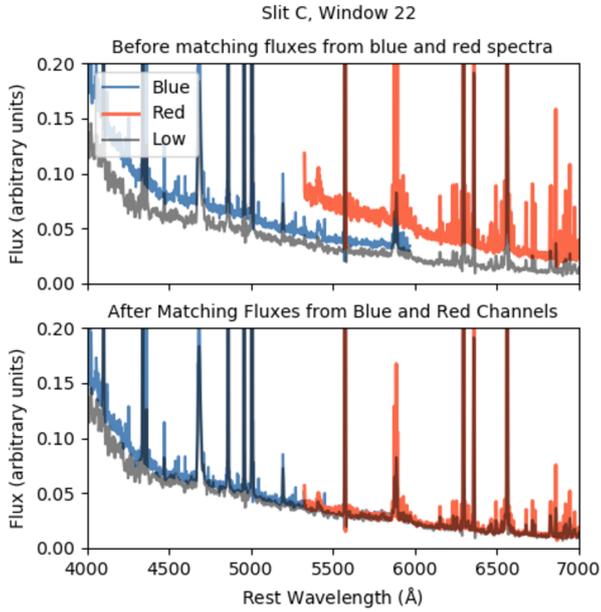

**Figure B1.** Spectra from the blue channel (thin blue curve) and red channel (thick red curve) before and after matching the flux levels (top and bottom panels respectively). The grey curve shows the low dispersion spectrum which was used for matching both channels.